\documentclass{elsart}

\usepackage{graphicx}
\usepackage{amssymb}

\begin{document}

\begin{frontmatter}

\title{QCD Dirac spectrum and components of the gauge field}

\author{Arata~Yamamoto}
\ead{a-yamamoto@ruby.scphys.kyoto-u.ac.jp}
\address{Department of Physics, Faculty of Science, Kyoto University, \\Kitashirakawa, Sakyo, Kyoto 606-8502, Japan}

\begin{abstract}
We analyze the relation between the Dirac spectrum and the gauge field in SU(3) lattice QCD.
We focus on how a certain component of the gauge field is related to the Dirac spectrum.
First, we consider momentum components of the gauge field.
It turns out that the broad momentum region is relevant for the low-lying Dirac spectrum and topological charges.
The connection with chiral random matrix theory is also discussed.
Second, we consider an SU(2) subgroup component of the SU(3) gauge field.
The SU(2) subgroup component behaves like the SU(2) gauge field in the low-lying Dirac spectrum.
\end{abstract}

\begin{keyword}
Lattice QCD \sep Dirac spectrum \sep Random matrix theory
\PACS 11.15.Ha \sep 12.38.-t \sep 12.38.Aw \sep 12.38.Gc
\end{keyword}
\end{frontmatter}

\section{Introduction}
The eigenvalue spectrum of the Dirac operator is a fundamental quantity in QCD.
It reflects important properties induced by the gauge field.
In particular, zero modes of the Dirac operator have special roles on nonperturbative properties of QCD.
The number of zero modes determines the topological charge $Q$ of the background gauge field, as
\begin{eqnarray}
n_R - n_L = N_f Q,
\label{eqQF}
\end{eqnarray}
where $n_R$ and $n_L$  are the numbers of right-handed and left-handed zero modes, respectively \cite{At68}.
Also, the existence of zero modes is related to spontaneous chiral symmetry breaking, through the Banks-Casher relation,
\begin{eqnarray}
 \pi \rho(0) = \Sigma ,
\end{eqnarray}
where $\rho(0)$ is the macroscopic spectral density of zero modes and $\Sigma$ is the absolute value of the chiral condensate \cite{Ba80}.

Moreover, interestingly, the low-lying Dirac spectrum shows the universal behavior of disordered systems.
Such an universal behavior is successfully described by chiral random matrix theory \cite{Sh93,Gu98,Ve00}.
The distribution of the low-lying Dirac spectrum depends only on the global symmetries of the system, and not on the details of the microscopic structure.
Once the validity of chiral random matrix theory is guaranteed, chiral random matrix theory enables us to predict the low-lying eigenvalue distribution by a simple and universal function.

It is definitely important to clarify how the gauge field induces these properties of the Dirac spectrum.
However, it is rather difficult.
These properties are beyond the reach of perturbation theory.
Although the Dirac equation is written in a simple form, the low-energy Dirac spectrum is not easily obtained because of the complicated dynamics of the gauge field.
In chiral random matrix theory, the information about the dynamics of the gauge field is lost.

The aim of this study is to investigate the relation between the Dirac spectrum and the gauge field in lattice QCD.
We focus on two kinds of components of the gauge field: momentum components and subgroup components.
We decompose the gauge field into some components, and then calculate the Dirac spectrum from the decomposed components.
From the change of the Dirac spectrum, we analyze how the components of the gauge field are related to the Dirac spectrum.
Although such a decomposition is gauge dependent, we can intuitively understand the role of the gauge field from this analysis.

In this paper, we calculate the staggered Dirac spectrum in lattice QCD.
The basics of the staggered Dirac spectrum and the details of the numerical simulations are shown in section 2.
For the analysis of momentum components, we introduce a lattice framework of momentum cutoff in section 3.
We discuss the resultant Dirac spectrum and the connection with chiral random matrix theory.
For the analysis of subgroup components, we project the SU(3) gauge field onto an SU(2) subgroup component, as demonstrated in section 4.
Finally, section 5 is devoted to the summary.

\section{The Dirac spectrum in lattice QCD}
\subsection{Staggered Dirac operator}
The Euclidean Dirac operator is defined as
\begin{eqnarray}
D = \sum_\mu [ \gamma_\mu \partial_\mu - ig\gamma_\mu A_\mu(x) ].
\end{eqnarray}
The $j$-th eigenmode of the Dirac operator is given by
\begin{eqnarray}
D \psi_j = i \lambda_j \psi_j.
\end{eqnarray}
The chirality is defined as
\begin{eqnarray}
\chi_j = \psi^\dagger_j \gamma_5 \psi_j.
\end{eqnarray}
In continuum QCD, the anti-commutation relation,
\begin{eqnarray}
\{ D, \gamma_5 \} = 0,
\label{eqAC}
\end{eqnarray}
is satisfied.
Because the Dirac operator is anti-Hermitian, $\lambda_j$ is zero or nonzero real.
Nonzero eigenvalues are always paired as $\pm i\lambda_j$ and their chiralities are zero.
On the other hand, zero eigenvalues are unpaired, and their chiralities are $+1$ or $-1$.

The staggered Dirac operator is
\begin{eqnarray}
D = \frac{1}{2} \sum_\mu \eta_\mu (x) [\bar{U}_\mu(x) \delta_{x+\hat{\mu},y} - \bar{U}^\dagger_\mu(x-\hat{\mu})  \delta_{x-\hat{\mu},y} ],
\label{eqDst}
\end{eqnarray}
where $\eta_\mu (x)$ is the standard staggered phase factor.
For the standard (unimproved) staggered Dirac operator, $\bar{U}_\mu(x)$ is the SU(3) link variable $U_\mu(x)$.
For the improved staggered Dirac operator, $\bar{U}_\mu(x)$ is a combination of the link variables.

The staggered fermion describes the four-flavor Dirac fermion in continuum limit.
At finite lattice spacing, however, there is an $O(a^2)$ term which violates the flavor symmetry and Eq.~(\ref{eqAC}).
Due to the presence of this term, the staggered Dirac spectrum does not have exact zero eigenvalues, unlike the continuum Dirac spectrum.
As a result, the unimproved staggered Dirac operator fails to reproduce nontrivial topological sectors \cite{Go99,Da99a,Da99b}.
This problem is approximately settled by improving the gauge and fermion actions \cite{Du04,Fo04a,Wo05,Fo05}.

\subsection{Simulation details}
We computed the low-lying eigenvalues of the staggered Dirac operator by SU(3) quenched lattice QCD simulations \cite{Go99,Da99a,Da99b,Du04,Fo04a,Wo05,Fo05}.
We used the improved gauge action and the improved staggered Dirac operator.

\begin{table}[b]
\begin{center}
\renewcommand{\tabcolsep}{0.5pc} 
\renewcommand{\arraystretch}{1} 
\caption{\label{tab1}
The parameters of quenched gauge configurations.
The lattice coupling $\beta$, the corresponding lattice spacing $a$, the lattice volume $V$, and the configuration number $N_{\rm conf}$ are listed.
}
\begin{tabular}{ccccc}
\hline
$\beta$ & $a$ (fm) & $V$ ($a^4$) & $V$ (fm$^4$) & $N_{\rm conf}$ \\
\hline
9.0 & 0.07 & $20^4$ & $1.4^4$ & 50 \\
8.6 & 0.08 & $16^4$ & $1.3^4$ & 50 \\
8.3 & 0.11 & $12^4$ & $1.3^4$ & 50 \\
7.9 & 0.16 & $8^4$  & $1.3^4$ & 5000 \\
\hline
\end{tabular}
\end{center}
\end{table}

For the gauge action, we adopted the 1-loop improved Symanzik gauge action with tadpole improvement.
The 1-loop improved Symanzik gauge action includes the plaquette term, the $1\times 2$ rectangle term, and the $1\times 1 \times 1$ ``parallelogram'' term \cite{Lu85}.
The tadpole coefficient is defined as the fourth-root of the average plaquette value \cite{Le93}.
As listed in Table \ref{tab1}, we generated four kinds of quenched gauge configurations.
The physical volumes of these gauge configurations are approximately the same.
The $12^4$, $16^4$, and $20^4$ lattices are used for the analysis of eigenvalues and topological charges.
The $8^4$ lattice is used for the discussion about chiral random matrix theory.

For the staggered Dirac operator, we adopted the FAT7$\times$ASQ operator \cite{Fo04b}.
The FAT7$\times$ASQ operator is constructed from the FAT7 link variable and the ASQ operator \cite{Or99}.
This improved operator strongly suppresses the flavor-symmetry breaking lattice artifact.
The spatial boundary conditions are periodic, and the temporal boundary condition is antiperiodic.
Because the operator is anti-Hermitian, the eigenvalues always appear in pairs, $\pm i\lambda_j$.
We only have to calculate the positive eigenvalues $\lambda_j > 0$.
It should be understood that the negative eigenvalues with the same magnitude exist.
We calculated the low-lying 30 positive eigenvalues.

\begin{figure}[t]
\begin{center}
\includegraphics[scale=1]{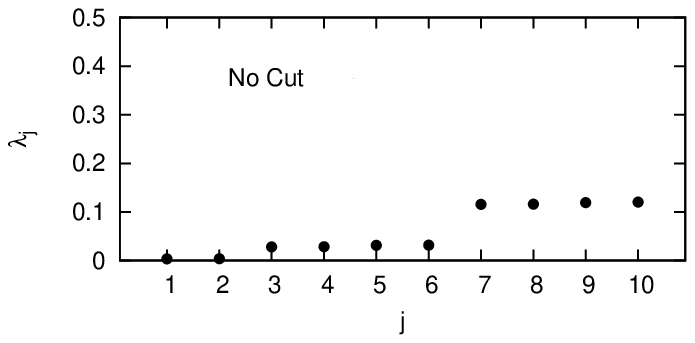}
\caption{\label{fig1}
The eigenvalue spectrum of the FAT7$\times$ASQ Dirac operator.
Only low-lying 10 positive eigenvalues of one typical gauge configuration are shown.
The calculation is done on the $16^4$ lattice with $\beta=8.6$.
}
\end{center}
\begin{center}
\includegraphics[scale=1.1]{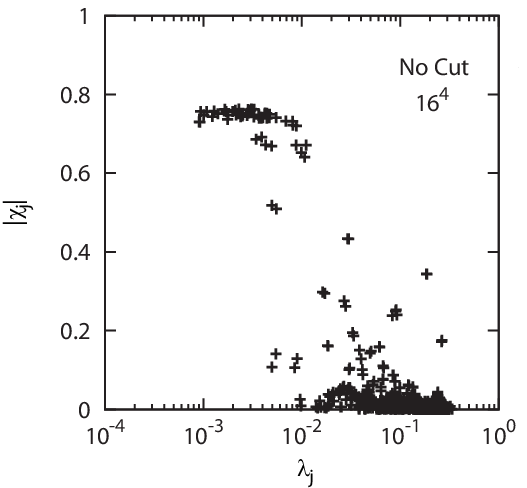}
\caption{\label{fig2}
The scatter plot of the eigenvalue $\lambda_j$ and the absolute value of the chirality $\chi_j$.
The low-lying 30 eigenvalues of 50 gauge configurations are plotted.
The calculation is done on the $16^4$ lattice with $\beta=8.6$.
}
\end{center}
\end{figure}

In Fig.~\ref{fig1}, we show the low-lying 10 positive eigenvalues of one typical gauge configuration of $\beta=8.6$.
The four-fold degeneracy, i.e., the four-flavor symmetry, is satisfied with high precision.
In addition, zero eigenvalues are approximately reproduced.
Such zero modes are called as ``would-be'' or ``near'' zero modes.
Because the negative eigenvalues with the same magnitudes exist, the topological charge of this gauge configuration is $Q=1$.
In Fig.~\ref{fig2}, we show the scatter plot of all the low-lying 30 eigenvalues of 50 gauge configurations of $\beta=8.6$.
There exists a clear separation between the nonzero modes, which is $|\chi_j|\simeq 0$, and the would-be zero modes, which is $|\chi_j|\simeq 0.8$.
The reason why the chirality deviates from $\pm 1$ is that it requires a renormalization \cite{Sm88}.

\section{Analysis of momentum components}
In this section, we analyze how the Dirac spectrum is changed when the gauge field is restricted to some momentum components.
Since the behavior of the gauge field depends on the energy scale, different momentum components would have different roles. 
In a different point of view, this analysis relates to the connection between the energy scales of the fermion field and the gauge field.

\subsection{Momentum cutoff}
Here, we briefly introduce the lattice framework to remove some momentum components of the gauge field \cite{Ya08,Ya09}.
The framework is formulated as the following five steps.

Step 1. The link variable $U_{\mu}(x)$ is generated by Monte Carlo simulation.
The link variable is fixed with a certain gauge.

Step 2. The momentum-space link variable ${\tilde U}_{\mu}(p)$ is obtained by the Fourier transformation, as
\begin{eqnarray}
{\tilde U}_{\mu}(p)=\frac{1}{V}\sum_{x} U_{\mu}(x)\exp(i \sum_{\nu} p_\nu x_\nu ),
\end{eqnarray}
where $V$ is the lattice volume.

\begin{figure}[t]
\begin{center}
\includegraphics[scale=0.45]{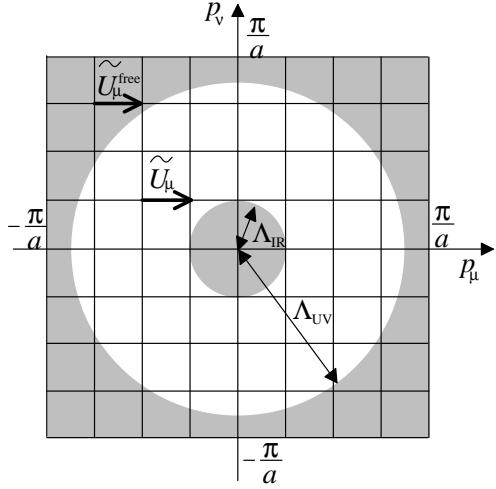}
\caption{\label{fig3}
The schematic figure of momentum space.
}
\end{center}
\end{figure}

Step 3. Some components of ${\tilde U}_{\mu}(p)$  are removed by introducing a momentum cutoff.
In the cutoff region, the momentum component is replaced by the free gauge field $A_\mu=0$, i.e., the free link variable
\begin{equation}
{\tilde U}^{\rm free}_{\mu}(p)=\frac{1}{V}\sum_x 1 \exp(i {\textstyle \sum_\nu} p_\nu x_\nu)=\delta_{p0}.
\end{equation}
For example, as one of the most natural choices, the ultraviolet cutoff $\Lambda_{\rm UV}$ is defined by 
\begin{equation}
\label{eqUV}
{\tilde U}_{\mu}^{\Lambda}(p)= \Bigg\{
\begin{array}{cc}
{\tilde U}_{\mu}(p) & (\sqrt{p^2} \le \Lambda_{\rm UV})\\
0 & (\sqrt{p^2} > \Lambda_{\rm UV}),
\end{array}
\end{equation}
and the infrared cutoff $\Lambda_{\rm IR}$ is defined by
\begin{equation}
\label{eqIR}
{\tilde U}_{\mu}^{\Lambda}(p)= \Bigg\{
\begin{array}{cc}
\delta_{p0} & (\sqrt{p^2} < \Lambda_{\rm IR})\\
{\tilde U}_{\mu}(p) & (\sqrt{p^2} \ge \Lambda_{\rm IR}).
\end{array}
\end{equation}
The schematic figure is shown in Fig.~\ref{fig3}.

Step 4. The coordinate-space link variable is obtained by the inverse Fourier transformation as
\begin{eqnarray}
U'_{\mu}(x)=\sum_{p} {\tilde U}_{\mu}^{\Lambda}(p)\exp (-i \sum_{\nu} p_\nu x_\nu ).
\end{eqnarray}
Since $U'_{\mu}(x)$ is not an SU(3) matrix in general, $U'_{\mu}(x)$ must be projected onto an SU(3) element $U^{\Lambda}_{\mu}(x)$.
The projection is realized by maximizing the quantity
\begin{eqnarray}
{\rm ReTr}[\{ U^{\Lambda}_{\mu}(x) \}^{\dagger}U'_{\mu}(x)].
\end{eqnarray}

Step 5. The expectation value of an operator $O$ is computed by using this link variable $U^{\Lambda}_{\mu}(x)$ instead of $U_{\mu}(x)$, i.e., $\langle O[U^\Lambda]\rangle$ instead of $\langle O[U]\rangle$.

Using this framework, we analyze how the physical quantity is changed by the momentum cutoff.
From the resultant change, we can nonperturbatively investigate the relation between the physical quantity and momentum components of the gauge field.

While the Dirac spectrum is gauge invariant, a momentum component of the gauge field is gauge variant.
This is because a gauge transformation is nonlocal in momentum
space, and some momentum component could become another momentum 
component by a gauge transformation. 
Thus, we must fix the link variable with some gauge in Step 1.
Although we can choose any gauge, we mainly used the Landau gauge in this paper.

\subsection{Eigenvalue and chirality}
\begin{figure}[t]
\begin{center}
\includegraphics[scale=1]{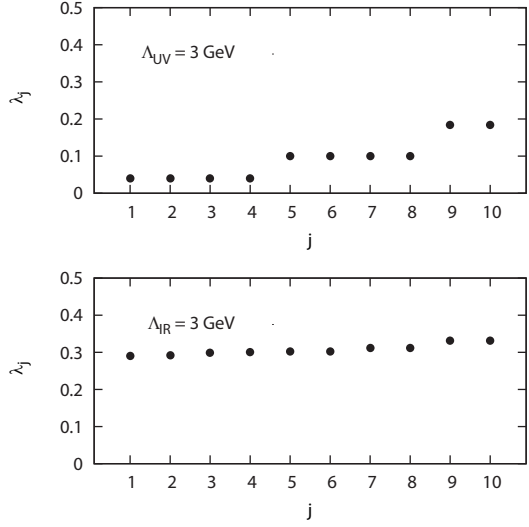}
\caption{\label{fig4}
The low-lying Dirac spectrum with the momentum cutoff.
The original gauge configuration is the same as that of Fig.~\ref{fig1}.
}
\end{center}
\end{figure}

\begin{figure}[t]
\begin{center}
\includegraphics[scale=0.8]{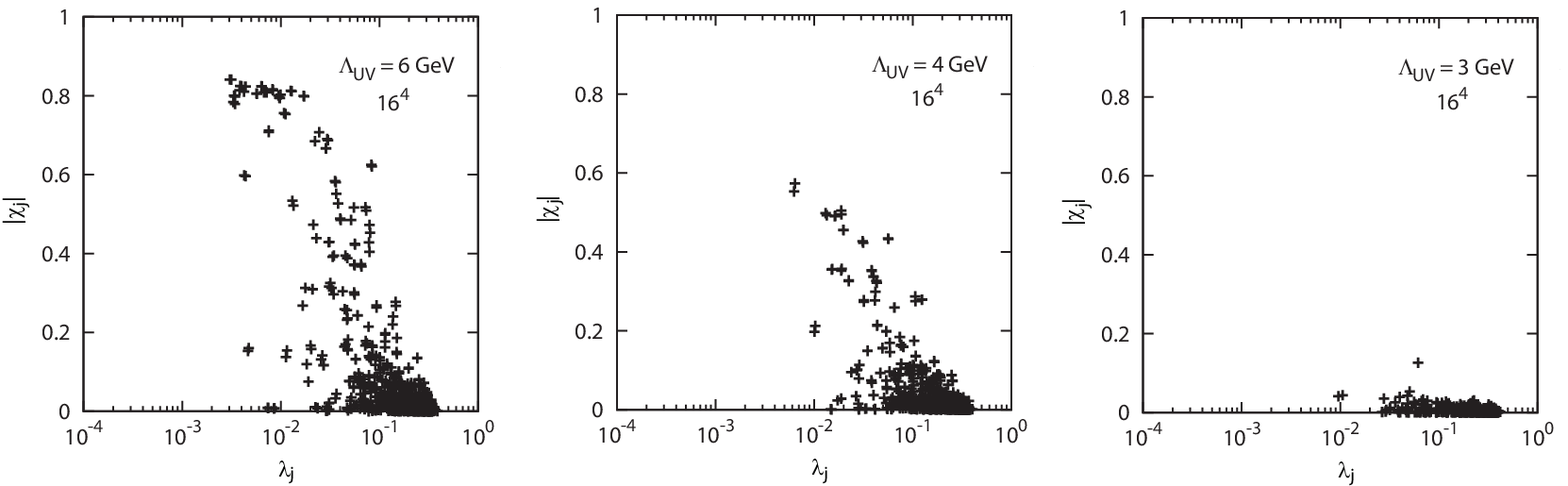}
\caption{\label{fig5}
The scatter plot of the eigenvalue $\lambda_j$ and the absolute value of the chirality $\chi_j$ with the ultraviolet cutoff $\Lambda_{\rm UV}$.
The low-lying 30 eigenvalues of 50 gauge configurations are plotted.
The calculation is done on the $16^4$ lattice with $\beta=8.6$.
}
\end{center}
\begin{center}
\includegraphics[scale=0.8]{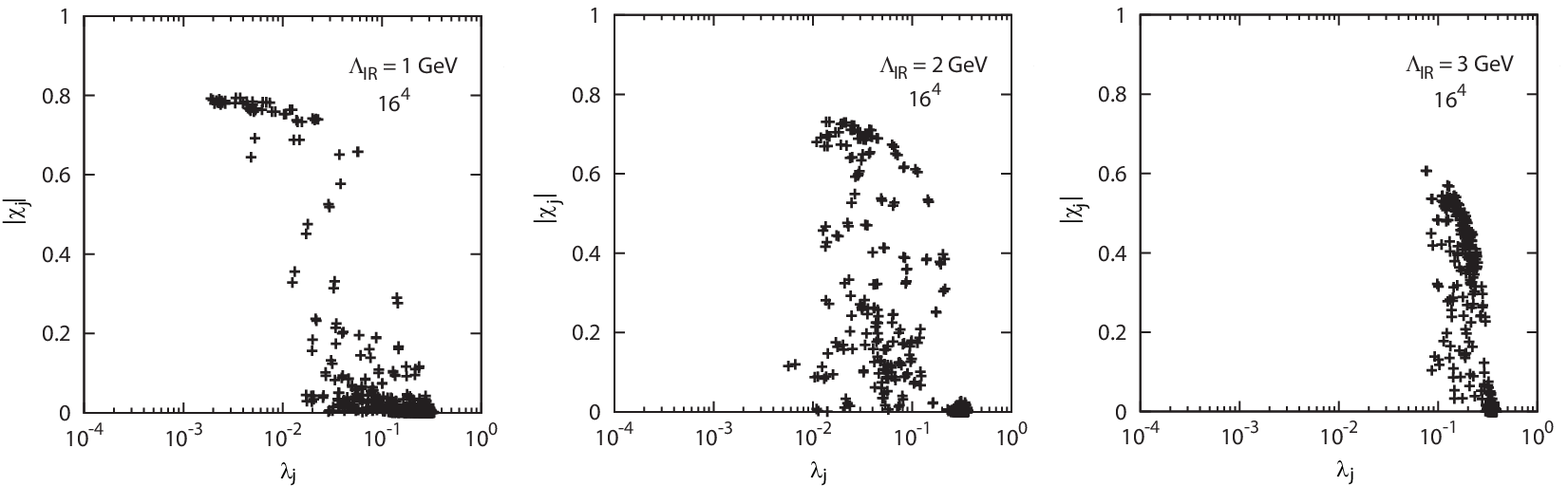}
\caption{\label{fig6}
The scatter plot of the eigenvalue $\lambda_j$ and the absolute value of the chirality $\chi_j$ with the infrared cutoff $\Lambda_{\rm IR}$.
The low-lying 30 eigenvalues of 50 gauge configurations are plotted.
The calculation is done on the $16^4$ lattice with $\beta=8.6$.
}
\end{center}
\end{figure}

We applied the above framework to the Dirac spectrum.
In Fig.~\ref{fig4}, we show the Dirac spectrum with $\Lambda_{\rm UV}=3$ GeV and $\Lambda_{\rm IR}=3$ GeV.
The original gauge configuration is the same as that of Fig.~\ref{fig1}.
For $\Lambda_{\rm UV}=3$ GeV, while all the eigenvalues change, the four-fold degeneracy is still satisfied.
The would-be zero modes which exist in Fig.~\ref{fig1} disappear.
This means that the topological charge is broken.
For $\Lambda_{\rm IR}=3$ GeV, the low-lying 10 eigenvalues become close to the eigenvalue of free fermions, which is about 0.39 for this case.
Also in this case, the topological charge becomes zero.

In Fig.~\ref{fig5}, we show the scatter plot of the eigenvalue and the chirality with the ultraviolet cutoff $\Lambda_{\rm UV}$.
As the high-momentum component is removed, the eigenvalue spectrum is gradually changed.
For $\Lambda_{\rm UV}=6$ GeV, the distribution is not so different from the original one.
For $\Lambda_{\rm UV}=3$ GeV, all eigenmodes are $|\chi_j|\simeq 0$.
Therefore, all gauge configurations fall into the trivial topological sector.

The scatter plot with the infrared cutoff $\Lambda_{\rm IR}$ is shown in Fig.~\ref{fig6}.
When the low-momentum component is removed, many of the low-lying eigenvalues approach the free eigenvalue, which is $\lambda_j \simeq 0.39$ and $|\chi_j|\simeq 0$.
Compared to the case of the ultraviolet cutoff, the changes of the eigenvalues are sensitive to the infrared cutoff.
This result is consistent with our expectation that the low-lying Dirac eigenmode interacts with the low-momentum gauge field.
For $\Lambda_{\rm IR}=3$ GeV, we cannot see would-be zero modes since all eigenvalues are almost degenerate.

\begin{figure}[t]
\begin{center}
\includegraphics[scale=0.8]{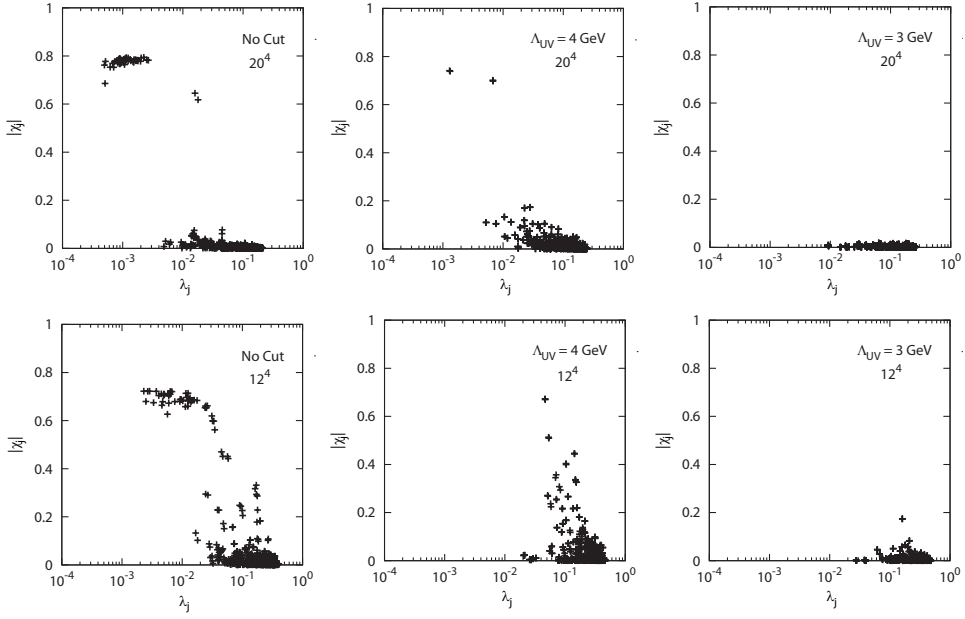}
\caption{\label{fig7}
The scatter plot of the eigenvalue $\lambda_j$ and the absolute value of the chirality $\chi_j$ with the ultraviolet cutoff $\Lambda_{\rm UV}$.
The results on the $20^4$ and $12^4$ lattices are plotted.
}
\end{center}
\end{figure}

To examine discretization effects, we also calculated the $12^4$ lattice with $\beta = 8.3$ and the $20^4$ lattice with $\beta = 9.0$.
The results are summarized in Fig. \ref{fig7}.
The distribution of the eigenvalue and the chirality depends on lattice spacing; the would-be zero modes are more clearly identified in the case of finer lattice spacing.
This behavior has been known in the standard lattice QCD \cite{Wo05,Fo05}.
In contrast, the existence of zero modes and its dependence on the momentum cutoff are insensitive to lattice spacing; the would-be zero modes are gradually destroyed at $\Lambda_{\rm UV}=4$ GeV, and then all eigenmodes become nonzero modes at $\Lambda_{\rm UV}=3$ GeV.
The lattice regularization is one kind of the ultraviolet cutoff, i.e., momentum is limited to less than a half width of the Brillouin zone $\pi/a$, as shown in Fig.~\ref{fig3}.
When lattice spacing is fine enough, the extra-high-momentum component near $\pi/a$ gives rise to renormalization effects, but does not contribute to physical observables.
Actually, this extra-high-momentum component does not affect the existence of zero modes.

When the momentum cutoff is introduced, the clear separation between would-be zero modes and nonzero modes is lost.
This is because the staggered Dirac operator does not have exact zero modes.
This situation will be improved by the overlap operator, which has exact zero modes \cite{Ne98,Ad10}.

\subsection{Topological charge}
From the number of would-be zero modes, we extracted the topological charge $Q$ of the background gauge field.
The topological charge is given through Eq.~(\ref{eqQF}).
We set the criterion for would-be zero modes as $|\chi_j|> 0.6$, and for nonzero modes as $|\chi_j|< 0.6$.
This criterion clearly separates would-be zero modes in the original case, as shown in Fig.~\ref{fig2}.
For the cases in which the clear separation between would-be zero modes and nonzero modes is lost, this criterion could be subtle.
The results are summarized in Table \ref{tab2}.

For a consistency check, we also extracted the topological charge in another way.
We directly calculated the topological charge by
\begin{eqnarray}
Q=\frac{1}{32\pi^2}\int d^4x \sum_{\mu\nu\sigma\rho} \epsilon_{\mu\nu\sigma\rho} {\rm Tr}[ F_{\mu\nu}(x) F_{\sigma\rho}(x) ].
\end{eqnarray}
Its discretized form is
\begin{eqnarray}
Q=-\frac{1}{512\pi^2}\sum_x \sum_{\mu\nu\sigma\rho=\pm 1}^{\pm 4} \epsilon_{\mu\nu\sigma\rho} {\rm Tr}[ U_{\mu\nu}(x) U_{\sigma\rho}(x) ],
\label{eqQG}
\end{eqnarray}
where $U_{\mu\nu}(x)$ is the plaquette.
We measured this topological charge after the APE smearing steps \cite{Da99b,Al87}.
The smearing coefficient and the number of the smearing step are set to 7.0 and 200, respectively.
The results are shown in the parentheses in Table \ref{tab2}.
The obtained topological charge is roughly consistent with the estimate by the would-be zero modes, especially in the original lattice QCD (``No Cut''), as expected.
These two estimates deviate from each other in some cases with the momentum cutoff, e.g., $\Lambda_{\rm UV}=4$ GeV.
This is partly because it is difficult to identify the would-be zero modes in these cases, and partly because the APE smearing steps change the local topological structure.

\begin{table}[t]
\begin{center}
\renewcommand{\tabcolsep}{0.5pc} 
\renewcommand{\arraystretch}{1} 
\caption{\label{tab2}
Topological charge $Q$ estimated by would-be zero modes.
The geometrical estimate by Eq.~(\ref{eqQG}) is shown in the parenthesis.
The results of the 50 gauge configurations are shown.
}
\begin{tabular}{cccccc}
\hline
$V$ ($a^4$) & Data & $|Q|=0$ & $|Q|=1$ & $|Q|=2$ & $|Q|=3$\\
\hline
$16^4$ &No Cut & 28 (29) & 18 (17) & 4 (4) & 0 (0) \\
&$\Lambda_{\rm UV}=6$ GeV & 35 (27) & 12 (17) & 3 (6) & 0 (0) \\
&$\Lambda_{\rm UV}=4$ GeV & 50 (44) &  0 (6)  & 0 (0) & 0 (0) \\
&$\Lambda_{\rm UV}=3$ GeV & 50 (50) &  0 (0)  & 0 (0) & 0 (0) \\
&$\Lambda_{\rm IR}=1$ GeV & 29 (27) & 17 (19) & 4 (4) & 0 (0) \\
&$\Lambda_{\rm IR}=2$ GeV & 32 (43) & 13 (7)  & 5 (0) & 0 (0) \\
&$\Lambda_{\rm IR}=3$ GeV & 49 (50) &  1 (0)  & 0 (0) & 0 (0) \\
\hline
$20^4$ &No Cut & 28 (28) & 22 (22) & 0 (0) & 0 (0) \\
&$\Lambda_{\rm UV}=4$ GeV & 50 (50) &  0 (0)  & 0 (0) & 0 (0) \\
&$\Lambda_{\rm UV}=3$ GeV & 50 (50) &  0 (0)  & 0 (0) & 0 (0) \\
\hline
$12^4$ &No Cut & 28 (27) & 20 (19) & 2 (2) & 0 (0) \\
&$\Lambda_{\rm UV}=4$ GeV & 49 (44) &  1 (6)  & 0 (0) & 0 (0) \\
&$\Lambda_{\rm UV}=3$ GeV & 50 (50) &  0 (0)  & 0 (0) & 0 (0) \\
\hline
\end{tabular}
\end{center}
\end{table}

\begin{table}
\begin{center}
\renewcommand{\tabcolsep}{0.5pc} 
\renewcommand{\arraystretch}{1} 
\caption{\label{tab3}
Topological charge $Q$ in the Coulomb gauge.
The results of the 50 gauge configurations of the $16^4$ lattice are shown.
}
\begin{tabular}{cccccc}
\hline
$V$ ($a^4$) & Data & $|Q|=0$ & $|Q|=1$ & $|Q|=2$ & $|Q|=3$\\
\hline
$16^4$ &No Cut & 28 (29) & 18 (17) & 4 (4) & 0 (0) \\
&$\Lambda_{\rm IR}=1$ GeV & 29 (27) & 17 (20) & 4 (3) & 0 (0) \\
&$\Lambda_{\rm IR}=2$ GeV & 35 (31) & 13 (14) & 1 (5) & 1 (0) \\
&$\Lambda_{\rm IR}=3$ GeV & 49 (50) &  1 (0)  & 0 (0) & 0 (0) \\
\hline
\end{tabular}
\end{center}
\end{table}

We can see that the number of the topological charges is gradually reduced in the both cases of the ultraviolet and infrared cutoffs.
The topological charge is almost completely destroyed at $\Lambda_{\rm IR}=3$ GeV or $\Lambda_{\rm UV}=3$ GeV.
This means that the topological charge can be destroyed by removing the momentum region to some extent, regardless of whether it is ultraviolet or infrared.
In other words, there is no typical energy scale of the gauge field inducing the topological charge.
This is different from the case of color confinement, which is induced only by the narrow low-momentum component of the gauge field \cite{Ya08,Ya09,YaS}.

We can study any other gauges in the same manner, instead of the Landau gauge.
In Table \ref{tab3}, we show the numerical results with the Coulomb gauge.
The results are almost the same as those of the Landau gauge.

\subsection{Connection with chiral random matrix theory}
Next, we compare the lattice data with a prediction of chiral random matrix theory, using the gauge configurations of $\beta=7.9$ and $V=8^4$.
This lattice volume is large enough to discuss the connection with chiral random matrix theory, i.e., in the so-called $\varepsilon$-regime \cite{Wo05}.
Here, we analyze the $Q=0$ sector with the criterion for nonzero modes as $|\chi_j|< 0.1$.

\begin{figure}[t]
\begin{center}
\includegraphics[scale=1]{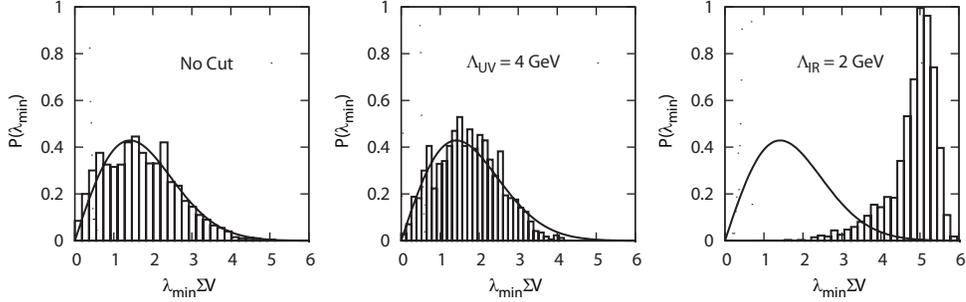}
\caption{\label{fig8}
The probability distribution of the lowest eigenvalue $\lambda_{\rm min}$ in the $Q=0$ sector.
The histograms are the lattice data of the $8^4$ lattice with $\beta=7.9$.
The solid curves for No Cut and $\Lambda_{\rm UV}=4$ GeV are the best-fit functions of Eq.~(\ref{eqP}) from chiral random matrix theory.
}
\end{center}
\end{figure}

We calculated the probability distribution of the lowest nonzero eigenvalue $\lambda_{\rm min}$.
In chiral random matrix theory, the probability distribution of the lowest eigenvalue is given as
\begin{eqnarray}
P(\lambda_{\rm min}) &=& \frac{z_{\rm min}}{2} e^{-z_{\rm min}^2/4}
\label{eqP}\\
z_{\rm min} &\equiv& \lambda_{\rm min} \Sigma V
\end{eqnarray}
for the SU(3) quenched gauge field with $Q=0$ \cite{Fo93}.
If one directly calculates the chiral condensate in infinite volume, this function is parameter free.
In this study, we treated $\Sigma$ as a fitting parameter.
The lattice data and the best-fit result of Eq.~(\ref{eqP}) are shown in Fig.~\ref{fig8}.
In the original lattice QCD (No Cut), the lattice data is well reproduced by chiral random matrix theory.
The best-fit value is $\Sigma \simeq 0.0022\times a^{-3}$.

For $\Lambda_{\rm UV}=4$ GeV, the qualitative behavior does not change.
The lattice data is still reproduced by Eq.~(\ref{eqP}).
Thus, the high-momentum component of the gauge field is irrelevant for the universality of the lowest eigenvalue distribution.
The best-fit value of $\Sigma$ is changed by the ultraviolet cutoff.
This is because the chiral condensate is renormalization-group variant and ultraviolet divergent.
By definition, the ultraviolet divergent quantity depends on the ultraviolet cutoff, although this might be irrelevant for phenomenology.

For $\Lambda_{\rm IR}=2$ GeV, the lowest eigenvalue distribution becomes to concentrate on the vicinity of the free eigenvalue.
The lattice data drastically deviates from Eq.~(\ref{eqP}).
Therefore, the low-momentum component of the gauge field is crucial for the validity of chiral random matrix theory.
In other words, the low-momentum component induces the strong-interacting and disordered nature of the gauge field.

In a previous work, the chiral condensate was directly analyzed by the lattice framework of the momentum cutoff \cite{Ya10}.
It suggests that the zero-momentum component of the gauge field has a large contribution to the chiral condensate.
To check it in the present study, we calculated the Dirac spectrum with the small infrared cutoff $\Lambda_{\rm IR}\sim 0.1$ GeV.
For $\Lambda_{\rm IR}\sim 0.1$ GeV, the lattice data can be fitted by Eq.~(\ref{eqP}) with the best-fit value $\Sigma \simeq 0.0019\times a^{-3}$.
The contribution of the zero-momentum component is indeed large, about 15\% of the total.
This is qualitatively consistent with the previous work.
Note, however, that this is quantitatively different from the previous work because of many systematic differences.

\section{Analysis of subgroup components}
Although there are many similarities between the SU(2) and SU(3) gauge theories, there are several differences because of an additional symmetry of SU(2).
An SU(3) group element includes SU(2) subgroup elements.
In this section, we consider whether an SU(2) subgroup component of the SU(3) gauge field shows the same behavior as the SU(2) gauge field.

\subsection{Subgroup projection}
The framework of the subgroup projection is formulated as below.

Step 1. The link variable $U_{\mu}(x)$ is generated by Monte Carlo simulation.
The link variable is fixed with a certain gauge.

Step 2. 
From the SU(3) link variable,
\begin{eqnarray}
U_{\mu}(x) = \left(
\begin{array}{ccc}
U_{11} & U_{12} & U_{13} \\
U_{21} & U_{22} & U_{23} \\
U_{31} & U_{32} & U_{33}
\end{array}
\right),
\end{eqnarray}
the SU(2)-projected link variable is generated from its subgroup element, as
\begin{eqnarray}
u_{\mu}(x) = \left(
\begin{array}{ccc}
U_{11} & U_{12} &0\\
U_{21} & U_{22} &0\\
0&0&1
\end{array}
\right).
\end{eqnarray}

Step 3. The expectation value of an operator $O$ is computed by using this link variable $u_{\mu}(x)$ instead of $U_{\mu}(x)$, i.e., $\langle O[u]\rangle$ instead of $\langle O[U]\rangle$.

\subsection{Numerical results}
The numerical analysis was performed in the same way as the previous section.
Also for the subgroup projection, we used the Landau gauge.
We can expect that other possible embeddings of the SU(2) subgroup give rise to the same results because the matrix elements are uniform in the Landau gauge.

\begin{figure}[t]
\begin{center}
\includegraphics[scale=1]{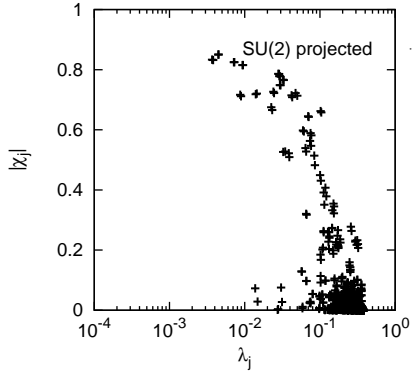}
\caption{\label{fig9}
The scatter plot of the eigenvalue $\lambda_j$ and the absolute value of the chirality $\chi_j$ of the SU(2)-projected gauge field.
The low-lying 20 eigenvalues of 50 gauge configurations are plotted.
The calculation is done on the $16^4$ lattice with $\beta=8.6$.
}
\end{center}
\end{figure}

We show the scatter plot of the eigenvalues and the chiralities in Fig.~\ref{fig9}, and the list of the topological charges in Table \ref{tab4}.
As for would-be zero modes and topological charges, the SU(2)-projected gauge field is qualitatively similar to the original SU(3) gauge field.
The Dirac spectrum has would-be zero modes, and the SU(2) topological charge exists.
Note that the number of would-be zero modes in each gauge configuration is different from that of the original SU(3) theory.

\begin{table}[t]
\begin{center}
\renewcommand{\tabcolsep}{0.5pc} 
\renewcommand{\arraystretch}{1} 
\caption{\label{tab4}
Topological charge $Q$ estimated by would-be zero modes.
The geometrical estimate by Eq.~(\ref{eqQG}) is shown in the parenthesis.
The results of the 50 gauge configurations with $\beta=8.6$ are shown.
}
\begin{tabular}{cccccc}
\hline
$V$ ($a^4$) & Data & $|Q|=0$ & $|Q|=1$ & $|Q|=2$ & $|Q|=3$\\
\hline
$16^4$ & SU(3) & 28 (29) & 18 (17) & 4 (4) & 0 (0) \\
& SU(2) projected & 38 (35) & 10 (11) & 2 (3) & 0 (1) \\
\hline
\end{tabular}
\end{center}
\end{table}

\begin{figure}[t]
\begin{center}
\includegraphics[scale=1.3]{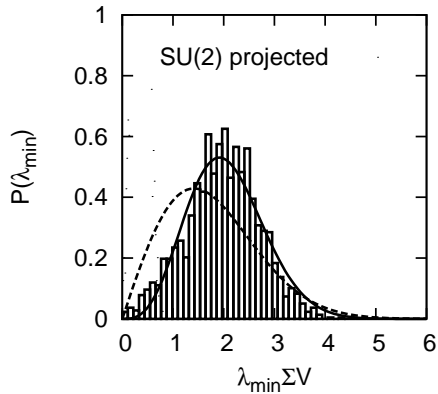}
\caption{\label{fig10}
The lowest eigenvalue distribution of the SU(2)-projected gauge field in the $Q=0$ sector.
The solid and dashed curves are Eqs.~(\ref{eqPSU2}) and (\ref{eqP}), respectively.
}
\end{center}
\end{figure}

The low-lying Dirac spectrum depends on the symmetry of the fermion field \cite{Ve94}.
The eigenvalue distributions are completely different between SU(3) and SU(2).
For the staggered Dirac operator with the SU(2) quenched gauge field, chiral random matrix theory predicts that the lowest eigenvalue distribution is
\begin{eqnarray}
P(\lambda_{\rm min}) &=& (z_{\rm min} \cosh z_{\rm min} - \sinh z_{\rm min} ) e^{-z_{\rm min}^2/2}
\label{eqPSU2}
\end{eqnarray}
in the $Q=0$ sector \cite{Fo93}.
This distribution is indeed obtained in SU(2) lattice QCD \cite{Be98}.
In Fig.~\ref{fig10}, we show the lattice data of the SU(2)-projected gauge field.
The solid and dashed curves are the predictions of chiral random matrix theory for SU(2) and SU(3), i.e., Eqs.~(\ref{eqPSU2}) and (\ref{eqP}), respectively.
The lattice data is reproduced by the prediction for SU(2), not for SU(3).
Therefore, the SU(2) subgroup component of the SU(3) gauge theory shows the same universal behavior as the SU(2) gauge theory.
The best-fit value of the chiral condensate is $\Sigma \simeq 0.0013\times a^{-3}$.
The ratio of this SU(2)-projected chiral condensate to the original SU(3) chiral condensate is 0.6, which is roughly understood as the ratio of the numbers of color.

\section{Summary}
We investigated the relation between the Dirac spectrum and components of the gauge field.
We considered momentum components and a subgroup component of the gauge group.
As for momentum components of the gauge field, the Dirac zero mode and the topological charge are gradually destroyed by the ultraviolet or infrared cutoff.
The low-momentum component is especially crucial for the validity of chiral random matrix theory.
As for an SU(2) subgroup component of the SU(3) gauge group, it behaves like the SU(2) gauge field.

The relation between zero modes and the gauge field is relevant for the topological structure of QCD vacuum, e.g., the topological charge density distribution.
It would be instructive to perform the similar analysis with the overlap operator.

\section*{Acknowledgments}
The author is supported by a Grant-in-Aid for Scientific Research [(C) No.~20$\cdot$363].
This work was supported by the Global COE Program, ``The Next Generation of Physics, Spun from Universality and Emergence,'' at Kyoto University.
This work was in part based on the MILC collaboration's public lattice gauge theory code (http://physics.utah.edu/\~{}detar/milc.html).
The lattice QCD simulations were carried out on Altix3700 BX2 and SX8 at YITP in Kyoto University.

\end{document}